\documentclass[aps,pre,twocolumn,superscriptaddress,groupedaddress]{revtex4}

\usepackage[utf8]{inputenc}
\usepackage[english]{babel}
\usepackage{indentfirst}
\usepackage{fancyhdr}
\usepackage[final]{showkeys}
\usepackage{graphicx}
\usepackage{subfig}
\usepackage{amssymb}
\usepackage{latexsym}
\usepackage{amsthm}
\usepackage{amsmath}
\usepackage{eucal}
\usepackage{amsfonts}
\usepackage{mathtools}
\usepackage{pgf,tikz}
\usepackage{mathrsfs}
\usepackage{epstopdf}
\usetikzlibrary{arrows}
\usepackage{enumerate}
\usepackage{epigraph}
\usepackage[a4paper, left=1.5cm, right=1.5cm, top=2cm, bottom=2cm]{geometry}
\usepackage{floatpag}

\DeclarePairedDelimiter\bra{\langle}{\rvert}
\DeclarePairedDelimiter\ket{\lvert}{\rangle}
\DeclarePairedDelimiterX\braket[2]{\langle}{\rangle}{#1 \delimsize\vert #2}
\DeclarePairedDelimiterX\inner[2]{\langle}{\rangle}{#1,#2}
\DeclarePairedDelimiterX\ketbra[2]{\lvert}{\rvert}{#1 \delimsize\rangle\! \delimsize\langle #2}

\newcommand{\me}[1]{\left\langle #1 \right\rangle }

\begin{document}

\title{Thermalization processes induced by quantum monitoring in multi-level systems}

\author{S. Gherardini}
\email{gherardini@lens.unifi.it}
\affiliation{SISSA, via Bonomea 265, I-34136 Trieste, \& INFN, Italy}
\affiliation{Department of Physics and Astronomy \& LENS, University of Florence, via G. Sansone 1, I-50019 Sesto Fiorentino, Italy}
\affiliation{CNR-IOM DEMOCRITOS Simulation Center, Via Bonomea 265, I-34136 Trieste, Italy}

\author{G. Giachetti}
\email{ggiachet@sissa.it}
\affiliation{SISSA, via Bonomea 265, I-34136 Trieste, \& INFN, Italy}

\author{S. Ruffo}
\email{ruffo@sissa.it}
\affiliation{SISSA, via Bonomea 265, I-34136 Trieste, \& INFN, Italy}
\affiliation{Istituto dei Sistemi Complessi, Consiglio Nazionale delle Ricerche, via Madonna del Piano 10, I-50019 Sesto Fiorentino, Italy}

\author{A. Trombettoni}
\email{andreatr@sissa.it}
\affiliation{Department of Physics, University of Trieste, Strada Costiera 11, I-34151 Trieste, Italy}
\affiliation{SISSA, via Bonomea 265, I-34136 Trieste, \& INFN, Italy}
\affiliation{CNR-IOM DEMOCRITOS Simulation Center, Via Bonomea 265, I-34136 Trieste, Italy}

\begin{abstract}
We study the heat statistics of a multi-level $N$-dimensional quantum system monitored by a sequence of projective measurements. The late-time, asymptotic properties of the heat characteristic function are analyzed in the thermodynamic limit of a high, ideally infinite, number $M$ of measurements $(M \to \infty)$. In this context, the conditions allowing for an Infinite-Temperature Thermalization (ITT), induced by the repeated monitoring of the quantum system, are discussed. We show that ITT is identified by the fixed point of a symmetric random matrix that models the stochastic process originated by the sequence of measurements. Such fixed point is independent on the non-equilibrium evolution of the system and its initial state. Exceptions to ITT, to which we refer to as \textit{partial thermalization}, take place when the observable of the intermediate measurements is commuting (or quasi-commuting) with the Hamiltonian of the quantum system, or when the time interval between measurements is smaller or comparable with the system energy scale (quantum Zeno regime). Results on the limit of infinite-dimensional Hilbert spaces ($N \to \infty$), describing continuous systems with a discrete spectrum, are also presented. We show that the order of the limits $M\to\infty$ and $N\to\infty$ matters: when $N$ is fixed and $M$ diverges, then ITT occurs. In the opposite case, the system becomes classical, so that the measurements are no longer effective in changing the state of the system. A non trivial result is obtained fixing $M/N^2$ where instead partial ITT occurs. Finally, an example of partial thermalization applicable to rotating two-dimensional gases is presented.
\end{abstract}

\maketitle

\section{Introduction}

Quantum monitoring refers in general to the action of performing a sequence of quantum measurements on a system or a portion of it\,\cite{JacobsContempPhys2006,WisemanBook,WeberNature2014,RossiPRL2020}. Being the single quantum measurement a dynamical process with probabilistic nature, it is customary to associate to any sequence of measurements a stochastic process obeying, over time, to a specific probability distribution\,\cite{WisemanBook,JacobsBook}. Such distribution usually depends on properties that rely on both the system and the measured observable, and also external sources of noise\,\cite{HatridgeScience2013,GherardiniPRA2019}.

The study of sequences of quantum measurements, especially projective ones, is present in several physical systems and applications, ranging from fundamental quantum physics and quantum Zeno phenomena\,\cite{ItanoPRA1990,KwiatPRL1995,KofmanNature2000,FischerPRL2001,FacchiJPA2008, WoltersPRA2013,ZhuPRL2014,SchaferNatComm2014,SignolesNatPhys2014,GherardiniNJP2016,ChaudhrySciRep2017}, quantum metrology and sensing\,\cite{KiilerichPRA2015,KohlhaasPRX2015,MuellerPRA2016,PiacentiniNatPhys2017,SchioppoNatPhot2017,DoNJP2019,SakuldeePRA2020,MuellerPLA2020} to
quantum thermodynamics\,\cite{CampisiPRL2010,CampisiPRE2011,HorowitzPRE2012,YiPRE2013,FuscoPRX2014,GherardiniQST2017,ElouardNPJ2017,BayatPRL2018,GherardiniPRE2018,GarciaPintoPRL2019,MartinsPRA2019,Hernandez2019,GiachettiCondMatt2020,Hernandez2021}, both at the theoretical and experimental level. In particular, protocols implementing repeated measurements have been already successfully applied to investigate the quantum Zeno effect/dynamics\,\cite{SchaferNatComm2014,SignolesNatPhys2014,GherardiniQST2017}. Instead, in quantum metrology repeated measurements can be used to probe the phase evolution of an atomic ensemble thanks to interleaved interrogations and feedback corrections, see for example\,\cite{KohlhaasPRX2015,SchioppoNatPhot2017} and, recently, also to carry out quantum noise sensing, as shown e.g.\,in\,\cite{DoNJP2019,SakuldeePRA2020,MuellerPLA2020}. In addition, an active line of research focuses on the characterization of the thermodynamics principles that rule the statistics of the measurement outcomes, with several contributions making use of quantum fluctuation theorems and Jarzynski relations\,\cite{EspositoRMP2009,CampisiRMP2011,SagawaBook2013,JaramilloPRE2017,DenzlerPRE2018,GherardiniQST2018,ManzanoPRX2018}. Within this framework, since each measurement entails a sudden energy variation with a given probability, one can also analyze the probability distribution of the heat exchanged by a monitored quantum system with its surroundings, as done in Refs.\,\cite{GherardiniPRE2018,GiachettiCondMatt2020} for two and three-level quantum systems. Moreover, also the monitoring of local observables in quantum many-body systems have been recently investigated\,\cite{NationPRE2020,RossiniPRB2020}. Specifically, in \cite{NationPRE2020} it has been observed that the measurement outcomes of a macroscopic observable may evolve by following a Brownian diffusion dynamics, while in \cite{RossiniPRB2020} the interplay between unitary Hamiltonian driving and random local projective measurements is analyzed at the quantum transition point of a quantum lattice spin systems, by showing that local measurement processes generally tend to suppress quantum correlations.

In this paper, we study the asymptotic behaviour of a $N$-level quantum system subjected to a randomly distributed sequence of quantum projective measurements. As figure of merit, we consider the statistics of the heat distribution exchanged by the system with its surroundings. Our main motivation is three-fold. {\em (i)} There is an inherent difference in the response of a quantum system to a sequence of projective measurements depending whether it has a finite number of levels (say $N$) or it is continuous; thus, we aim at investigating how the limit of large $N$ affects the results found for finite $N$ such as the ones presented in\,\cite{GherardiniPRE2018,GiachettiCondMatt2020}. {\em (ii)} For spin-$s$ systems, the classical limit is retrieved for $s \to \infty$, so a natural question is to study how the effects of quantum measurements change by varying/increasing the quantum spin label $s$ counting the possible projections $s_z$, whose number is $2s+1$ that plays the role of the number of levels $N$ [in the sense that the observables, including the ones measured in the monitoring process, are operators with dimension $(2s+1) \times (2s+1$)]. {\em (iii)} We are also motivated by recent experimental results obtained on negatively charged nitrogen-vacancy (NV) centers\,\cite{Hernandez2019}. An NV center is a localized impurity in diamond lattice based on a nitrogen substitutional atom and a nearby vacancy. In the NV experiment in\,\cite{Hernandez2019}, it has been possible to locally address the impurity and perform a sequence of quantum projective measurements along the $z$-axis, not commuting in such case with the energy eigenbasis of the system. In\,\cite{Hernandez2019} it has been observed a tendency of the quantum system towards an equilibrium thermal state with infinite temperature, which can be seen as an instance of an Infinite-Temperature Thermalization (ITT) process.

Similar behaviours can be also observed in periodically driven quantum systems, especially those used in Floquet engineering \cite{Eckardt2017,Oka2019}.
The reason for the analogy with the effect provided by repeated measurements as studied here is that the measurement apparatus could be seen as a periodic drive, and may transfer energy to the measured system, similarly to what the periodic drive may do. As discussed in\,\cite{GherardiniPRE2018}, it not very important whether the time intervals between
subsequent measurements are fixed or random obeying a certain distribution. In addition, for a periodically driven system with a convergent Magnus expansion, the drive allows for the system to relax towards a steady-state that is locally indistinguishable from the microcanonical ensemble of the Floquet Hamiltonian\,\cite{MoriJPB2018}. In particular, as argued in Refs.\,\cite{DalessioPRX2014,LazaridesPRE2014}, if a periodically driven quantum system is non-integrable, then it is expected to naturally evolve towards the infinite-temperature state, locally indistinguishable from almost all the other quantum states. Therefore, here, the natural arising question is what is the interplay between the number of levels of the analyzed system and the value associated to the independent parameters of the quantum monitoring protocol, i.e., the number of measurements and the time interval between them.

To our knowledge, in the literature there are no works that systematically discuss how internal energy fluctuations distribute over time in a $N$-level quantum system subjected to $M$ projective quantum measurements. Our paper aims at filling this gap, by predicting the non-equilibrium behaviour of the monitored system in the thermodynamic limits of $M$ and $N$ large, both ideally infinite. The projective measurements are defined by a generic Hermitian observable and separated by a not-zero time interval $\tau$. Note that, although we will mostly consider the case in which the time intervals $\tau$ are randomly chosen with average value $\tau$, the obtained results do not depend on the randomness in such time intervals.

The paper is structured as follows. In Sec.\,\ref{sec2} we describe the non-equilibrium dynamics to which a monitored $N$-level quantum system is subjected, while in Sec.\,\ref{sec:large_M_limit} the asymptotic behaviour of the quantum system dynamics, as well as of the its heat statistics, are analysed in the thermodynamics limit of a large (ideally infinite) number of intermediate projective measurements. In such a limit, ITT can occur. Exceptions to ITT are then addressed in Sec.\,\ref{sec:excep}, while in Sec.\,\ref{sec:spin_s_systems} our theoretical findings are tested on a spin-$s$ particle in a magnetic field. Then, in Sec.\,\ref{sec:large} we show results in the thermodynamic limit of $N$ large, and an example of partial thermalization in a rotating two-dimensional gas is discussed in Sec.\,\ref{sec:inc}. Finally, our conclusions are presented in Sec.\,\ref{sec:concl}.

\section{Non-equilibrium dynamics}\label{sec2}

Let us consider a quantum system defined in a $N$-dimensional Hilbert space whereby the Hamiltonian $H$, assumed to be time-independent, admits the following spectral decomposition
\begin{equation}
H = \sum^N_{k=1} E_k |E_k\rangle\!\langle E_k|.
\end{equation}
At time $t=0^{-}$ the system is supposed to be in an arbitrary quantum state described by the density operator $\rho_0$. We then apply the two-point measurement scheme\,\cite{TalknerPRE2007}, where a projective measurement of energy is performed both at the initial and at the final time of the protocol. Therefore, at time $t = 0^{+}$ a first projective energy measurement is carried out, with the result that the state of the system after the measurement is one of the projectors $|E_k\rangle\!\langle E_k|$ with probability $c_k$ (where $c_k >0 \ \forall k = 1, \dots, N$ and $\sum^N_{k=1} c_k =1$), while the energy of the system is $E_k$.

Afterwards, the system undergoes a number $M$ of consecutive projective measurements of the generic observable
\begin{equation}
\mathcal{O} \equiv \sum^N_{k=1} \alpha_k |\alpha_k\rangle\!\langle\alpha_k|
\end{equation}
where $\alpha_k$ and $\ket{\alpha_k}$ are the outcomes and eigenstates of $\mathcal{O}$, respectively. We suppose $[H,\mathcal{O}] \neq 0$.

The monitoring protocol is detailed as follows. Between the energy measurement at time $t = 0^{+}$ and the first measurement of $\mathcal{O}$, the system does not evolve apart from a trivial phase, since only the Hamiltonian acts in this time interval. After each measurement of $\mathcal{O}$ the state of the system is given by one of the projectors $|\alpha_k\rangle\!\langle\alpha_k|$ with probability $\pi_k = \operatorname{Tr}[\rho_0|\alpha_k\rangle\!\langle\alpha_k|]$\,\cite{vN}. During the time-interval between the $(j-1)^{\text{th}}$ and the $j^{\text{th}}$ measurement of $\mathcal{O}$, the system evolves according to the unitary dynamics generated by $H$, i.e., $U(\tau_j) = e^{-iH\tau_j}$, where $\hbar$ is set to unity and the waiting times $\tau_j$ denote the interval between two consecutive measurements. The latter may not be deterministic quantities, since also $\tau_j$ can be random variables distributed by following the joint Probability Density Function (PDF) $p(\tau_1,\dots,\tau_M)$. The numerical simulations in the considered cases show that taking the waiting times $\tau_j$ as random variables or fixed does not alter the results and the late-time dynamics of the system. The probability of finding the system in $|\alpha_k\rangle\!\langle\alpha_k|$ after the $M^{\text{th}}$ measurement is denoted as $\Tilde{\pi}_{k_M}$. Finally, a second energy measurement is performed immediately after the last, the $M^{\text{th}}$, measurement of $\mathcal{O}$. We denote by $E_m$ the outcome of the second and final energy measurement, whereby the final state of the system is $\ketbra{E_m}{E_m}$, and by $p_m$ the corresponding probability. It holds that $p_m = \sum_k \Tilde{\pi}_k |\braket{\alpha_k}{E_m}|^2$. Before proceeding, it is worth observing that the number of measurements $M$ and the waiting times $\tau_j$ depend each other through the relation $t_{\rm fin}=\sum_{j=1}^{M}\tau_{j}$, with $t_{\rm fin}$ denoting the final time of the monitoring protocol. Thus, $M$ and $\tau_j$ are \emph{independent} variables if we do not fix the value of $t_{\rm fin}$. This assumption will be maintained throughout the paper.

The variation of the system internal energy $\Delta U$ is defined as\,\cite{TalknerPRE2007}
\begin{equation}\label{eq:def_Q}
\Delta U \equiv E_m - E_n \,,
\end{equation}
which is thus a random variable. By considering each projective measurement as a random exogenous genuinely-quantum process, one can identify the internal energy variation $\Delta U$ as heat $Q$, absorbed or emitted by the system\,\cite{GherardiniPRE2018}.

In the following, we will denote by $\boldsymbol\tau \equiv (\tau_1, \dots,\tau_M)$ the sequence of waiting times and $\mathbf{k} \equiv (k_1, \dots, k_M)$ the sequence of the outcomes obtained by measuring $\mathcal{O}$ in the single protocol realization. As we are going to observe, the most important contribution to the variation of the system dynamics occurs during the application of the $M$ measurements of $\mathcal{O}$. For this purpose, let us introduce the conditional probability $P_{k_M|k_1}$ to get the outcome $\alpha_{k_M}$ from the $M^{\text{th}}$ measurement of $\mathcal{O}$, provided that the first intermediate-measurement outcome was $\alpha_{k_1}$. The conditional probability $P_{k_M|k_1}$ obeys the relation
\begin{equation}
    \Tilde{\pi}_{k_M} = \sum_{k_1} P_{k_M|k_1} \pi_{k_1}\,.
\end{equation}
Being all the $M$ measurements projective, one can check that
\begin{equation}\label{tildepi}
    P_{k_M|k_1} = \int d^{M} \boldsymbol\tau  \ p(\boldsymbol{\tau}) \sum_{k_{1}, \dots, k_{M-1}} \operatorname{Tr}\left[ \nu_{\mathbf{k},\boldsymbol\tau} |\alpha_{k_1}\rangle\!\langle\alpha_{k_1}| \nu^{\dagger}_{\mathbf{k},\boldsymbol\tau}\right]
\end{equation}
where we have introduced the quantities
\begin{equation}\label{Nu}
\begin{split}
\nu_{\mathbf{k},\boldsymbol \tau} &\equiv \ketbra{\alpha_{k_M}}{\alpha_{k_M}} U(\tau_{M-1}) \cdots \ketbra{\alpha_{k_2}}{\alpha_{k_2}} U(\tau_1)  \\
&= \prod^{M}_{j=3} \bra{\alpha_{k_{j}}} U(\tau_{j-1}) \ket{\alpha_{k_{j-1}}} \ketbra{\alpha_{k_M}}{\alpha_{k_2}} U(\tau_1).
\end{split}
\end{equation}
It is worth noting that Eq.\,\eqref{tildepi} can be rewritten,
in matrix notation, as:
\begin{equation} \label{matrixnotation}
    P_{k_M|k_1} = \int d^{M} \boldsymbol\tau \  p(\boldsymbol{\tau}) \bra{\alpha_{k_M}} \prod^M_{j=2} L(\tau_{j-1}) \ket{\alpha_{k_1}}
\end{equation}
with
\begin{equation}\label{eq:def_L_tau}
    \bra{\alpha_{k_{j-1}}} L(\tau_{j-1}) \ket{\alpha_{k_j}} \equiv \lvert \bra{\alpha_{k_{j-1}}} U(\tau_{j-1}) \ket{\alpha_{k_j}} \rvert^{2}.
\end{equation}

This expression has a clear physical interpretation in terms of the formalism of stochastic processes. As a matter of fact, the quantity
$|\bra{\alpha_{k_{j-1}}} U(\tau_{j-1}) \ket{\alpha_{k_j}}|^2$ is the conditional probability to obtain the outcome $\alpha_{k_j}$ from the $j^{\text{th}}$ projective measurement once measured the outcome $\alpha_{k_{j-1}}$ from the $(j-1)^{\text{th}}$ one. Then, each $L(\tau)$ can be seen as the \emph{transition matrix} pertaining to a discrete-time Markov chain in which the eigenstates of the observable $\mathcal{O}$ play the role of the states of the Markov chain. Consequently, the operator $L(\tau)$ is a \emph{stochastic matrix} with rows or columns summing to $1$. This property of $L(\tau)$ can be easily verified by observing
that
\begin{eqnarray}
\displaystyle{\sum_{k=1}^N \bra{\alpha_{\ell}} L(\tau) \ket{\alpha_k}} &=& \displaystyle{\sum_{k=1}^N \bra{\alpha_{\ell}} U(\tau) \ketbra{\alpha_k}{\alpha_k}  U^{\dagger}(\tau) \ket{\alpha_{\ell}}}\nonumber \\
&=&\bra{\alpha_{\ell}}U(\tau)U^{\dagger}(\tau)\ket{\alpha_{\ell}} = 1
\end{eqnarray}
$\forall\ell=1,\dots,N$.

This being said, the fluctuation profile of the heat $Q$ can be characterized by means of the characteristic function $G(u)$ (with $u\in\mathbb{C}$) associated to the probability distribution ${\rm P}(Q)$. By construction, the characteristic function is defined as
\begin{equation}
    G(u) \equiv \me{e^{iQu}} = \me{e^{i (E_m - E_n) u}}
\end{equation}
where the average $\langle\cdot\rangle$ is performed over a
large number of realizations of the underlying non-equilibrium dynamics.

In the following, the $M$-large behaviour of a monitored $N$-level quantum systems is analyzed by studying the asymptotic properties of the transition matrix $L(\tau)$ as well as the corresponding expression of $G(u)$.

\section{Infinite-temperature thermalization}\label{sec:large_M_limit}

In this paragraph, the asymptotic behaviour of $P_{k_M|k_1}$ is studied in the limit of $M \gg 1$. The time intervals $\tau_j$ are different from zero and on average greater than the energy scale of the analysed quantum system, for any $j=1,\ldots,M$. In this way, the system dynamics is not ``frozen'' as an effect of the quantum Zeno regime\,\cite{KofmanNature2000,FacchiPRL2002,FacchiJPA2008,SmerziPRL2012,SchaferNatComm2014,SignolesNatPhys2014,GherardiniNJP2016,MuellerAdP2017}. For a recent example studying the large-time dynamics of a many-body system (fermionic lattice) under the influence of a dephasing noice refer to \cite{RibeiroArxiv2020}, while an investigation of the convergence properties of the work distribution done by a quantum system when the number of its degrees of freedom (along regularized path integrals) goes to infinity is presented in Ref.\,\cite{vanZonPRE2008}.

Let us start observing that, being $\{L(\tau_j)\}_{j=1}^{M-1}$ transition matrices (expressed as a function of conditional probabilities), they are symmetric stochastic operators. In particular, since each element of the transition matrix $L(\tau_j)$ is the square modulus of the corresponding element of a unitary matrix (see Eq.\,(\ref{eq:def_L_tau})), the $L(\tau_j)$'s are unistochastic matrices. Thus, all its eigenvalues $\lambda_k$ are such that $|\lambda_k| \leq 1$ and at least one of them is equal to $1$. More formally, one can state that
$-1 \leq \lambda_k \leq 1$ with $k=1,\ldots,N$. For the sake of simplicity, we also assume that $\tau_1 = \dots = \tau_M \equiv \tau$. In the limit of large $M$, the product of the transition matrices $L(\tau)$ behaves asymptotically as a proper combination of the projectors $\mathcal{P}_{\lambda=1}$ and $\mathcal{P}_{\lambda=-1}$ associated, respectively, to the eigenspaces identified by $\lambda=1$ and $\lambda=-1$. In other terms,
\begin{equation}\label{eq:product_L}
L(\tau)^{M-1} \rightarrow \mathcal{P}_{\lambda=1} + (-)^{M-1} \mathcal{P}_{\lambda=-1}\,.
\end{equation}
However, while we are guaranteed that the eigenvalue $\lambda = 1$ actually exists for any $\tau$, the presence of the eigenvalue $\lambda=-1$ is not so obvious. For example, in the $N=2$ case, the smallest eigenvalue of $L$ is given by $\lambda = 1 - 2 \sin^2(\phi)\sin^2\left(\frac{\Delta E \tau}{2}\right)$, where $\Delta E$ denotes the energy gap of the qubit,
while $\phi$ is the angle that defines the rotation bringing the
eigenbasis of the Hamiltonian $H$ over the eigenbasis of the measurement observable $\mathcal{O}$. In order to get $\lambda = -1$, not only we need to choose a very specific value of $\mathcal{O}$ (i.e., an observable $\mathcal{O}$ such that $\sin(\phi) = \pm 1$), but we also need to assume $\tau^{\ast} = \frac{(2k+1)\pi}{\Delta E}$ with $k \in \mathbb{Z}$. It is clear that, apart from fine-tuned cases, the concurrence of both these conditions in a $N$-level system do not take place (especially if the time intervals $\tau_j$ are randomly distributed). As a result, one can expect on physical grounds that $\mathcal{P}_{\lambda=-1} = 0$ such that
\begin{equation}\label{bruttina}
L(\tau)^M \rightarrow \mathcal{P}_{\lambda=1} \,.
\end{equation}

However, it is important to note that Eq.\,\eqref{bruttina} does not imply that in the single realization of the system dynamics the
effects originated by the presence of rare fluctuations are absent. In such case, indeed, the evaluation of higher-order statistical moments
could be still required. For more details on the analysis of the impact of rare fluctuations in the statistics of quantum observables, the reader
can refer e.g.\,to Refs.\,\cite{GherardiniQST2017,GherardiniPRA2019} that analyze the problem by means of the large deviation theory.

What discussed so far holds for a generic stochastic matrix. However, being $L(\tau)$ also symmetric, one can verify that
\begin{equation}
\ket{v} = \frac{1}{\sqrt{N}}\sum_{k=1}^{N}\,\ket{\alpha_k}
\end{equation}
is such that $L(\tau) \ket{v} = \ket{v}$ for all values of $\tau$. This means that $\ket{v}$ is invariant to the application of the stochastic
matrix $L(\tau)$, or in other terms, $\ket{v}$ is a \emph{fixed point} of $L(\tau)$. If we assume that $\lambda=1$ is non degenerate, then $L(\tau)^{M-1} \rightarrow |v\rangle\!\langle v|$. Thus, since the eigevector $\ket{v}$ does not depend on the value of $\tau$, we can conclude that
\begin{equation} \label{bellina}
L(\tau_{M-1}) \cdots L(\tau_1) \rightarrow \ketbra{v}{v}
\end{equation}
also for randomly distributed $\tau$'s,
as long as the set of $\boldsymbol{\tau}$ for which $\lambda=-1$ is eigenvector, or $\lambda = 1$ is degenerate, has zero measure. However, such a degeneracy of $\lambda=1$ can occur and
the corresponding analysis is postponed to Sec.\,\ref{sub_A}. It is also worth noting that, in the Markov chain language, the validity of Eq.\,(\ref{bellina}) means that the underlying process is ergodic and admits a unique asymptotic configuration, i.e., the uniform one whereby the probabilities that the final state of the system is one of the eigenvectors $\ket{\alpha_k}$ of $\mathcal{O}$ are the same.

Let us explore the meaning of this property in our context. In the $M \gg 1$ limit, Eq.\,\eqref{matrixnotation} becomes
\begin{equation}
    P_{k_M|k_1} = \braket{\alpha_{k_M}}{v} \braket{v}{\alpha_{k_1}} = \frac{1}{N}
\end{equation}
so that, regardless from the state of the system after the first measurement of $\mathcal{O}$, one has:
\begin{equation}
    \Tilde{\pi}_{k_M} = \sum_{k_1} P_{k_M|k_1} \pi_{k_1} = \frac{1}{N}\,.
\end{equation}
Thus, as expected, the information on the initial condition is lost as $M$ increases. Moreover, this result is also independent on the form of the observable $\mathcal{O}$, and all the possible outcomes $|\alpha_{k_M}\rangle\!\langle\alpha_{k_M}|$ are equiprobable. Accordingly, the state of the system after the $M^{\text{th}}$ measurement (with $M \gg 1$) is described by the maximally mixed state
\begin{equation} \label{finalstate}
    \rho_M = \frac{\mathbb{I}}{N}\,.
\end{equation}
Note that, being $\rho_M$ diagonal in every basis, the second energy
measurement (corresponding to the last measurement of the whole non-equilibrium dynamics) has no effect, and also all the final energy
outcomes are equiprobable.
\begin{figure}
    \centering
    \subfloat[][]{\includegraphics[scale=0.575]{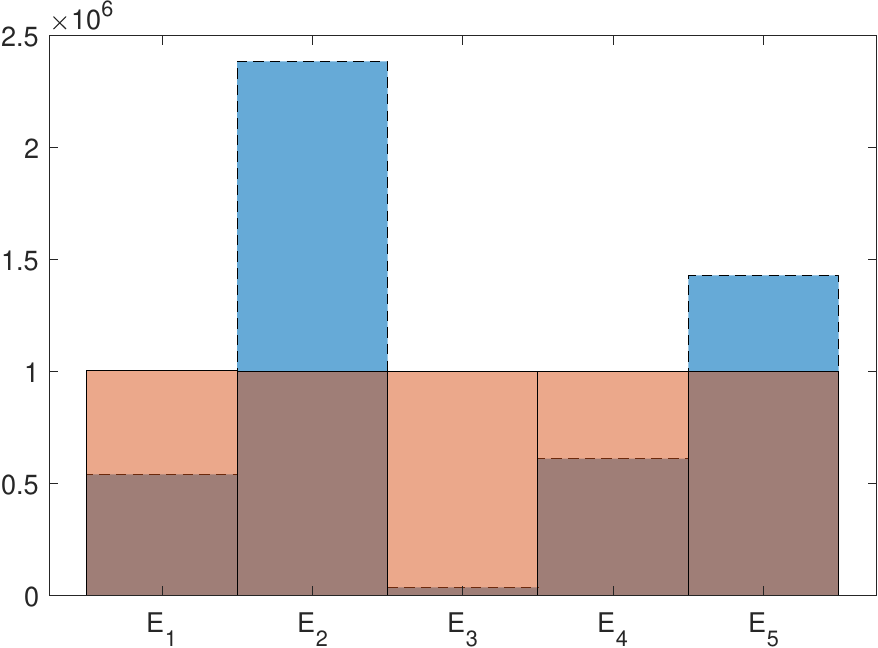}} \ \
    \subfloat[][]{\includegraphics[scale=0.575]{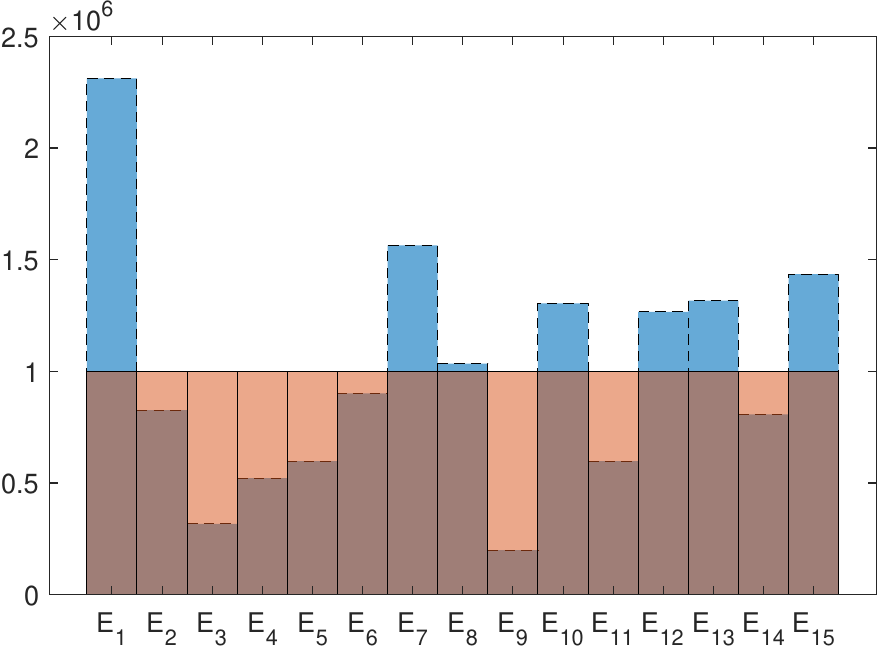}}
    \caption{Comparison between the initial (dashed-line histogram with blue-coloured area) and final (solid-line histogram with red-coloured area) heat statistics for a five-level $(a)$ and fifteen-level system $(b)$. The Hamiltonian of the system and the initial density operator $\rho_0$ are randomly chosen on a basis in which $\mathcal{O}$ is diagonal. The number of realizations of the non-equilibrium process is $5\cdot 10^6$ in $(a)$ and $15\cdot 10^6$ in $(b)$. In both cases, in the thermodynamic limit of $M$ large (in our numerical simulations $M=20$), each final energy value is equiprobable and such effect can be explained as the thermalization of the system towards a thermal state with $\beta=0$ (infinite temperature). In this figure and in the following ones, the parameter $\tau_j=1$ is chosen.}
    \label{fig:Hystogram}
\end{figure}

These findings are explicitly verified in Fig.\,\ref{fig:Hystogram},
where we plot for a $5$- and $15$-level quantum system the final energy
outcomes obtained at the end of the non-equilibrium dynamics of Sec.\,\ref{sec2}. Notice that in the figure $\tau_j=1$, but we verified that choosing $\tau_j$ as random variables (e.g., uniformly distributed) the final state at the end of the monitoring protocol is unaffected. The asymptotic behaviour occurring in the limit of $M$ large can be effectively interpreted as a thermalization process towards a thermal state with infinite temperature: $T=\infty$ ($\beta=0$). This can be understood by thinking that the measurement apparatus acts as a thermal reservoir with infinite energy (being it classical), by which, through a sequence of repeated interactions, a quantum system can reach the same equilibrium condition. In this respect, it is worth noting that the state of Eq.\,\eqref{finalstate} (maximally mixed state) maximizes the von Neumann entropy, and thus corresponds to the state associated to the absolute maximum of the entropy. For this reason, $\rho_M = \mathbb{I}/N$ has to be considered as the natural \emph{equilibrium state} for a quantum system to which no further constraints are imposed.

\subsection{Heat statistics}

As previously discussed, in the $M \rightarrow \infty$ limit the system ``forgets'' the initial state, meaning that it cannot be inferred by measurements of the system evolution. Thus, $E_m$ and $E_n$ are independent variables and $G(u)$ factorizes in the product of the characteristic functions of $E_n$ and $E_m$. The latter is given by
\begin{equation*}
    G_{E_m}(u) = \frac 1 N \sum^N_{n=1} e^{i u E_n} = \frac{1}{N} \operatorname{Tr} [e^{iuH}]\,,
\end{equation*}
since $\rho_M = \mathbb{I}/N$ and thus the values that $E_m$ can take
are uniformly distributed. Instead, the characteristic function of
$E_n$ equals to
\begin{equation*}
    G_{E_n}(u) = \sum_{k=1}^{N} \bra{\alpha_{k}} e^{-iHu} \rho_0 \ket{\alpha_{k}} = \operatorname{Tr}\left[e^{-iuH}\rho_0\right]
\end{equation*}
with the result that
\begin{equation}
G(u) = G_{E_n}(u) G_{E_m}(u) = \frac{1}{N}{\rm Tr}\left[e^{iHu}\right]{\rm Tr}\left[e^{-iHu}\rho_{0}\right].
\end{equation}
Consequently, by analyzing $G(u)$ at $u=i\epsilon$ with $\epsilon\in\mathbb{R}$, one gets
\begin{equation}\label{characteristic}
G(\epsilon) = \me{e^{- \epsilon Q}} = \frac{Z(\epsilon)}{N}\,{\rm Tr}\left[\rho_{0}\,e^{\epsilon H}\right],
\end{equation}
where $Z(\epsilon) \equiv {\rm Tr}[e^{-\epsilon H}]$ is the partition function of the Hamiltonian $H$ evaluated by taking $\epsilon$ as reference inverse temperature. As expected, if $\rho_0$ is a thermal state with inverse temperature $\epsilon = \beta$, we recover the
standard result $G(i\beta) = 1$, stemming directly from the Jarzynski equality.

\begin{figure}
    \centering
    \includegraphics[scale=0.58]{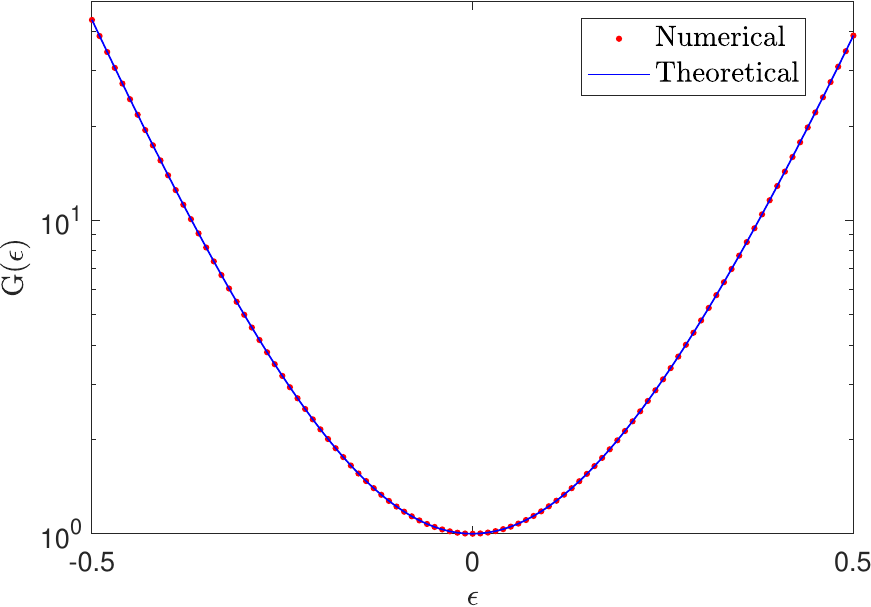}
    \caption{Comparison between the expression (\ref{characteristic})
    of the characteristic function $G(\epsilon) = \me{e^{-\epsilon Q}}$ (blue solid lines), plotted in semilogarithmic scale, and the numerical values (red dotted lines) computed for the fifteen-level system simulated in Fig.\,\ref{fig:Hystogram} panel (b).}
    \label{fig:Ge}
\end{figure}

In Fig.\,\ref{fig:Ge} we show the comparison between the results obtained by using Eq.\,(\ref{characteristic}) and the estimate of $G(u)$ from numerical simulations of the non-equilibrium process on the same fifteen-level systems used for Fig.\,\ref{fig:Hystogram} with $\rho_0$ random initial state. Excellent agreement is found.

Finally, from the knowledge of the characteristic function $G(u)$, we can also derive the statistical moments of the heat distribution. In doing this, let us compute the $n^{\rm th}$ derivative of $G(u)$ with respect to $u$, since
\begin{eqnarray}
\me{Q^n} &\equiv& (-i)^n \partial_u^n G(0)\nonumber \\
&=&\displaystyle{\sum_k \begin{pmatrix} n \\ k \end{pmatrix} (-1)^{n-k} \me{H^k}_{\infty}\me{H^{n-k}}_0}
\end{eqnarray}
where $\me{H^{\ell}}_{\infty} \equiv {\rm Tr}[\rho_{\infty}H^{\ell}]$ and $\me{H^{\ell}}_0 \equiv {\rm Tr}[\rho_{0}H^{\ell}]$, with $\ell$ integer number $\geq 1$. Note that, here, the subscripts $0$ and $\infty$ refer to the initial and asymptotic quantum states $\rho_0$ and $\rho_{\infty}$. Therefore, as expected, the first statistical moment $\me{Q}$ and the variance $\sigma^{2}(Q) \equiv \langle Q^2\rangle - \langle Q\rangle^2$ are respectively equal to
\begin{equation}
\begin{split}
\me{Q} &= \me{H}_{\infty} - \me{H}_0 \\
\sigma^2(Q) &= \sigma_{\infty}^{2}(H) + \sigma_{0}^{2}(H)\,.
\end{split}
\end{equation}
While the first moment $\me{Q}$ of the heat distribution depends on the sign of $\me{H}_{\infty}$ and $\me{H}_0$, in the large-$M$ limit the variance $\sigma^{2}(Q)$ is an additive function summing the variance of the initial and asymptotic energy distributions. Accordingly, thanks to this property, the variance of the heat distribution has to be preferred than the corresponding first moment to get information on the onset of thermalization in the limit of a large number of measurements.

\section{Partial thermalization}\label{sec:excep}

In the previous Sections, we have assumed that the largest eigenvalue $\lambda=1$ of $L(\tau)$ is non degenerate. Such assumption is realistic for a generic choice of the observable $\mathcal{O}$. However, interesting properties arise also if this assumption fails. Thus, in this paragraph we will analyze exceptions to Eq.\,(\ref{bellina}), leading to what we can refer to as \emph{partial} thermalization. Specifically, we will discuss the following cases: {\em (i)} $L(\tau_j)$ having a degenerate maximum eigenvalue; {\em (ii)} dynamics in the quantum Zeno regime; {\em (iii)} $[\mathcal{O},H]$ small. In the latter two cases, $L(\tau_j)$ is close to the identity matrix, so that the difference between the largest and the second-largest eigenvalues becomes small, allowing for a non-trivial interplay between the large number of measurements and the closing gap of the energy spectrum of the system.

\subsection{Eigenvalues degeneracy}\label{sub_A}

Let us assume that the largest eigenvalue $\lambda=1$ of $L(\tau)$ is degenerate. By construction, each element of $L(\tau)$ is $\geq 0$; thus, if $L(\tau)$ is a not reducible matrix (i.e., it cannot be put in a block diagonal form with a change of basis) the Perron-Frobenius theorem\,\cite{Perron} guarantees that the largest eigenvalue is non degenerate. Therefore, we have to consider the case in which $\mathcal{O}$
and $H$ share a common non trivial invariant subspace. This implies that
in the basis $\{\ket{\alpha_k}\}_{k=1}^{N}$, which defines the eigenstates
of $\mathcal{O}$, $H$ reads as
\begin{equation} \label{blockH}
H =  \begin{pmatrix} H_1 & &  \\ & \ddots & \\ & & H_R \end{pmatrix}
\end{equation}
where $R$ denotes the number of blocks of $H$ and $H_r$, with $r=1, \dots,R$,
are irreducible Hermitian matrices acting on the subspaces $S_r$. Before proceeding further, it is worth observing that having $H$ diagonal on the basis of $\mathcal{O}$ is a particular case of Eq.\,\eqref{blockH}, where each subspace has dimension one. From Eq.\,\eqref{blockH}, one can get that also the matrices $L(\tau_j)$ are block diagonal and can be written as
\begin{equation}\label{eq:L_partial_ITT}
L(\tau_j) =  \begin{pmatrix} L_1 (\tau_j) & &  \\ & \ddots & \\ & &  L_R (\tau_j) \end{pmatrix}
\end{equation}
where $L_r(\tau_j)$ are unistochastic irreducible matrices acting on the subspaces $S_r$ for $r=1,\ldots,R$ and $j=1,\ldots,M$. In this case, the Perron-Frobenius theorem ensures that no further degeneracy is present in each matrix $L_r(\tau_j)$. Therefore, we can introduce the set of eigenvectors, one for each subspace:
\begin{equation}
\ket{v_r} = \frac{1}{\sqrt{\dim{S_r}}} \sum_{k : \ket{\alpha_k} \in S_r} \ket{\alpha_k}
\end{equation}
corresponding to an $R$-order degeneracy of the eigenvalues of $L(\tau)$.
As a result, the eigenspace associated to the largest eigenvalue $\lambda=1$ is $R$ dimensional, and Eq.\,\eqref{finalstate} is no longer valid. Instead, one can find that $P_{k_M|k_1} = \frac{1}{\dim S_r}$ if $\ket{\alpha_{k_1}}$ and $\ket{\alpha_{k_M}}$ both belong to the same subspace $S_r$, and $P_{k_M|k_1} = 0$ otherwise. In such case, $\Tilde{\pi}_{k_M}$ keeps memory of the initial state. Indeed, if $\ket{\alpha_{k_M}} \in S_r$, then
\begin{equation}\label{eq:tilde_pi_partial_IIT}
    \Tilde{\pi}_{k_M}  = \frac{1}{\dim S_r} \sum_{k : \ket{\alpha_{k}} \in S_r} \pi_{k}\,.
\end{equation}
Since the initial and final energy projective measurements
does not mix the eigenspaces linked to the eigenvalues of $L(\tau)$, one can also write that
\begin{equation}\label{eq:pm_partial_IIT}
    p_m  = \frac{1}{\dim S_r} \sum_{k : \ket{E_{k}} \in S_r} c_k
\end{equation}
with $S_r$ such that $\ket{E_m} \in S_r$. In the case
of $R=N$ (namely $H$ commuting with $\mathcal{O}$: $[H,\mathcal{O}]=0$),
Eqs.\,(\ref{eq:tilde_pi_partial_IIT}) and (\ref{eq:pm_partial_IIT}) reduce,
as expected, to $\Tilde{\pi}_{k_M} = \pi_{k_1}$ and $p_m = c_m$, since in that case the evolution of the system is frozen and all the measurements outcomes coincide.

Moreover, by still assuming the degeneracy of $\lambda=1$, the heat characteristic function $G(u)$ can be written as the sum of the characteristic functions relative to each subspace $S_r$:
\begin{equation}\label{eq:G_partial_therm}
G(u) = \sum_{r=1}^{R} \frac{1}{\dim{S_r}}
{\rm Tr}\left[\rho_{\infty}\,e^{iH_{r}u}\right]{\rm Tr}\left[\rho_{0}\,e^{-iH_{r}u}\right] \,.
\end{equation}
From Eq.\,(\ref{eq:G_partial_therm}), it can be observed that also the moments of the heat distributions are provided by the sum of the corresponding moments for each subspace $S_r$.

These results have a simple physical interpretation.
For any realization of the introduced non-equilibrium process, after the first measurement, the state of the system is described by a vector belonging to $S_{\overline{r}}$ for some $\overline{r}\in\{1,\ldots,R\}$. Since such subspaces do not mix each other, the subsequent system evolution will take place within $S_{\overline{r}}$. As a result, in the limit of $M \rightarrow \infty$, the monitored quantum system tends to reach the completely mixed state in each $S_{r}$ separately.

An example of partial thermalization clearly showing this feature is presented in Sec.\,\ref{sec:inc}.

\subsection{Quantum Zeno regime}

Another possible exception to ITT can be observed when the value
of all the waiting times $\tau_j$, with $j=1,\ldots,M$, is on average
much smaller than the inverse of the energy scale of the
system\,\cite{KofmanNature2000,FacchiPRL2002,FacchiJPA2008,SmerziPRL2012,SchaferNatComm2014,SignolesNatPhys2014,GherardiniNJP2016,MuellerAdP2017}.
In particular, let us consider here the case in which the total
time $\sum_{j=1}^{M}\tau_j$ remains constant in the limit of large-$M$,
thus ensuring that each waiting time $\tau_j$ is infinitesimal. In this
limiting case, we expect to recover the quantum Zeno regime that prevents
the system to thermalize.

This effect can be shown by observing that in the quantum Zeno regime the
operators $U$ and $L$ are nearly close to the identity matrix. In particular,
\begin{equation}\label{eq:Zeno_1}
\bra{\alpha_k} U(\tau_j) \ket{\alpha_{\ell}} =\delta_{k,\ell} - i\tau_j \bra{\alpha_k}H\ket{\alpha_\ell} + O(\tau_j^2)
\end{equation}
so that
\begin{equation}\label{eq:Zeno_2}
\bra{\alpha_k} L(\tau_j) \ket{\alpha_\ell} = \delta_{k,\ell} + O(\tau_j^2).
\end{equation}
Since their sum is constant, in the large-$M$ limit all
the waiting times $\tau_j$, $j=1,\ldots,M$, go to zero as $M^{-1}$.
Thus, $O(\tau_j^2) = O(M^{-2})$ such that the conditional probability $P_{k_M|k_1}$ can be read as
\begin{equation}
\begin{split}
P_{k_M|k_1} &= \delta_{k_{1},k_{M}} + (M-1)O(M^{-2}) \\
&= \delta_{k_{1},k_{M}} + O(M^{-1}).
\end{split}
\end{equation}
This means that, in the limit $M \rightarrow \infty$, the system is frozen in one of the eigenstates of $\mathcal{O}$, in accordance with the quantum Zeno effect.

\subsection{$\mathcal{O}$ and $H$ quasi-commuting observables}

Here, let us examine the case in which $[H,\mathcal{O}]$ is small. Under this hypothesis, the eigenbases of both the observables are close to each other, and the unitary matrix $V$ with elements $V_{k,\ell} \equiv \braket{\alpha_k}{E_\ell}$ is close to the identity.
Being $V$ an unitary matrix, we are allowed to parametrize $V$ as $V=e^{iR\xi}$ with $R$ Hermitian operator normalized such that $\|R\|_{2}=1$ with $\|\cdot\|_{2}$ the usual $L^2$ norm. In our case, being $V \simeq \mathbb{I}$, the parameter $\xi$ is $\ll 1$. Moreover, by introducing the diagonal matrices $\Lambda(\tau_j) = {\rm diag}(e^{-iE_1 \tau_j},\ldots,e^{-iE_N \tau_j})$, the propagator $U(\tau_j)$ can be expressed in the $\mathcal{O}$ eigenbasis as
\begin{equation}
U(\tau_j) = V \Lambda(\tau_j) V^{\dagger} =  \Lambda \left( \mathbb{I} + i \xi (\Lambda^{\dagger} R \Lambda -R) + O(\xi^2) \right),
\end{equation}
or -- by components -- as
\begin{eqnarray}\label{quasidiag}
&\displaystyle{U_{k,\ell} (\tau_j)=}&\nonumber \\
&\displaystyle{e^{-iE_\ell \tau_j} \left(\delta_{k,\ell} + i \xi R_{k,\ell} (e^{(E_k-E_\ell) \tau_j}-1) + O(\xi^2) \right).}&
\end{eqnarray}
Accordingly, for $k \neq \ell$, the $(k,\ell)$-element of the transition matrix $L(\tau_j)$ equals to
\begin{eqnarray}
L_{k,\ell} (\tau_j) &\equiv& |U_{k,\ell}(\tau_j)|^2 \nonumber \\
&=& 4 \xi^2 |R_{k,\ell}|^2 \sin\frac{(E_k-E_\ell)\tau_j}{2} + O(\xi^3).
\end{eqnarray}
At variance, regarding the diagonal elements of $L(\tau_j)$, we do not actually need to compute them, since they are are fixed by the constraint $\sum_{k}L_{k,\ell}(\tau_j)=1$. This consideration is
quite useful, since the $O(\xi^2)$ terms in Eq.\,\eqref{quasidiag}, which we did not compute, would have given rise in $L_{k,k}(\tau_j)$ to $O(\xi^2)$-terms that cannot be neglected. In conclusion, the transition matrix $L(\tau_j)$ can be put in the following form:
\begin{equation}\label{eq:L_small_xi}
    L(\tau_j) = \mathbb{I} - \xi^2 \Delta (\tau_j) + O(\xi^3)
\end{equation}
where $\Delta$ is a real symmetric operator whose elements are given by
\begin{equation}
\begin{cases}
&\Delta_{k,\ell}(\tau_j) = - 4 |R_{k,\ell}|^2 \sin^2{\frac{(E_k-E_\ell) \tau_j}{2}}\,, \ \forall\,k \neq \ell \\
&\Delta_{k,k} (\tau_j) = - \displaystyle{\sum_{k \neq \ell}\Delta_{k,\ell}}\,.
\end{cases}
\end{equation}
By analysing Eq.\,(\ref{eq:L_small_xi}), one has that,
for any finite small value of $\xi\neq 0$, the system thermalizes if undergoes a non-equilibrium process composed by $M \gg \xi^{-2}$ projective measurements. In particular, by taking a measurement observable $\mathcal{O}$ allowing for a finite value of $\xi$, the system thermalizes in the limit $M \rightarrow \infty$, while by imposing from the beginning that $\xi \rightarrow 0$ one recovers the same findings observed in the quantum Zeno regime also in the large-$M$ limit. However, a non trivial result is obtained if the two limits are performed at the same time with the constraint $M \xi^2 = \widetilde{t}$. In this case, assuming for simplicity $\tau_j = \tau$ $\forall j=1,\ldots,M$, we find that
\begin{equation}\label{eq:finite_time_Euclidean_ev}
L(\tau)^M \rightarrow e^{- \Delta(\tau) \widetilde{t}} \, ,
\end{equation}
mimicking a finite-time Euclidean evolution with effective Hamiltonian $\Delta(\tau)$ for the effective time $\widetilde{t}$. Therefore,
\begin{equation}
    \Tilde{\pi}_{k_M}  =
   \sum_{k_1}\bra{\alpha_{k_M}}e^{- \Delta(\tau) \widetilde{t}} \ket{\alpha_{k_1}} \pi_{k_1}\,.
\end{equation}
Moreover, since the bases of $\mathcal{O}$ and $H$ coincide up to $O(\xi)$-terms, a similar relation also holds for the probability $p_m$ to measure the energy $E_m$ after the $2^{\text{nd}}$ energy measurement
of the process:
\begin{equation}
    p_m = \sum_{n}
    \bra{E_m}e^{- \Delta(\tau) \widetilde{t}}\ket{E_n}c_n \,.
\end{equation}
As final remark, we also observe that, by construction, the operator $\Delta(\tau)$ has always a zero mode, namely an eigenvector with vanishing eigenvalue. This entails that the ITT and quantum Zeno regimes are recovered in the limits $\widetilde{t} \rightarrow \infty$ and $\widetilde{t} \rightarrow 0$, respectively.

\section{Spin-s systems}\label{sec:spin_s_systems}

To test our theoretical findings, we consider a spin-$s$ particle in a magnetic field taken directed along the $z$-axis. In this case the quantum number $s$ play the role of $N$ since the observables are described by $(2s+1) \times (2s+1)$ matrices. Thus, the system Hamiltonian is $H = -\omega S_z$, whose spectrum (apart from a constant) is given by $E_k = \omega k$ with $k = 0,\dots,2s$.

\begin{figure}
    \centering
    \subfloat[][]{\includegraphics[scale=0.64]{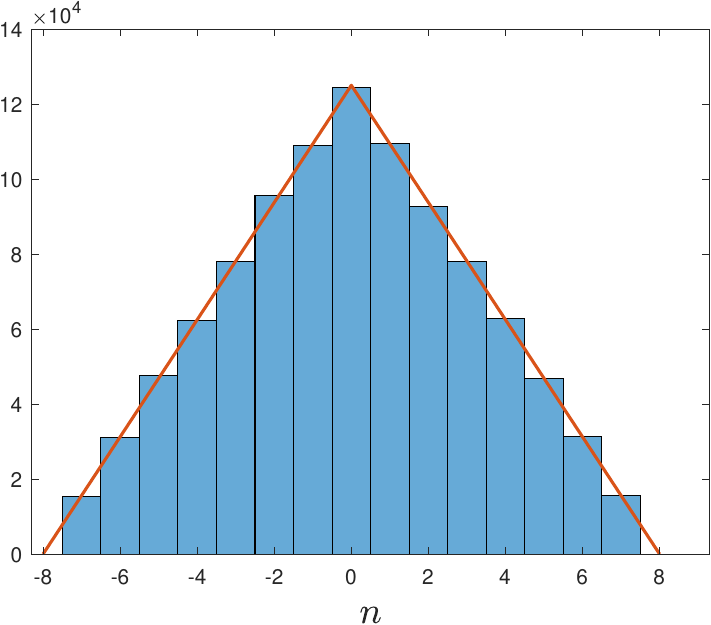}} \ \
    \subfloat[][]{\includegraphics[scale=0.64]{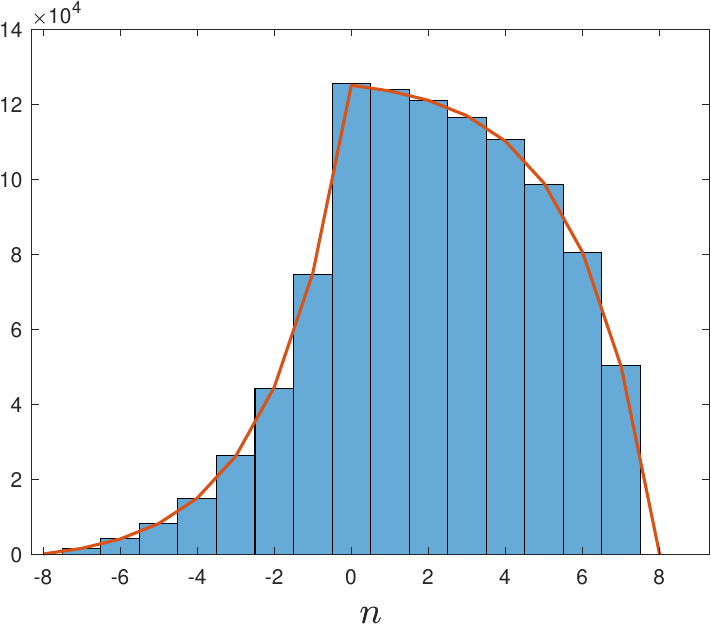}} \ \
    \subfloat[][]{\includegraphics[scale=0.64]{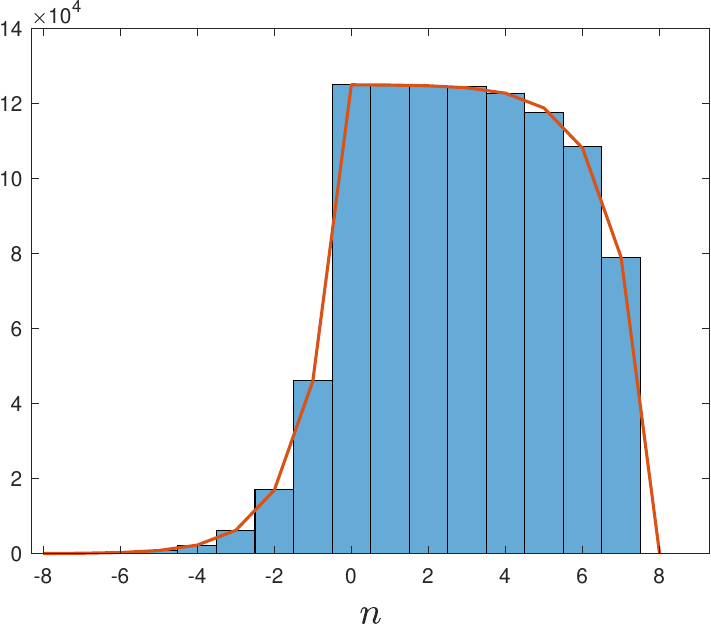}}
    \caption{Comparison between the theoretical estimate (red solid line) of the occurrence numbers to measure the heat outcomes $\omega\ell$, as provided by Eq.\,(\ref{spin}), and the corresponding histogram (blue areas). The latter has been obtained by numerically repeating the non-equilibrium dynamics of sequential measurements over $10^6$ realizations on a spin $s = \frac{7}{2}$. In the three panels, the initial state $\rho_0$ has been thermal with inverse temperature, respectively equal to $\beta = 0$, $\beta = 0.5$ and $\beta =1$.}
    \label{fig:spin}
\end{figure}

\subsection{Heat statistics}

Given a spin-$s$ particle in a magnetic field, let us assume that initial state $\rho_0$ of the spin is thermal, such that $c_k = e^{-\beta E_k}/Z$ with $Z={\rm Tr}[e^{-\beta H}]$ partition function. Under these assumptions, it is possible to compute exactly the probabilities associated to the heat distribution.

Being the energy levels of the spin evenly spaced, the outcomes of $Q$ are all the $4s+1$ values $Q = \omega\ell$ with $\ell = -2s,\dots,2s$. Since the spin operators $S_x$ and $S_y$ are non-commuting with $S_z$, if we choose to measure the spin component along these directions we will have ITT in the limit $M \gg 1$. Then, all the possible final outcomes $E_m$ will have the same probability $\frac{1}{2s+1}$ to occur. Hence, the probability $p_{\ell}(Q)$ to get the outcome $Q = \omega\ell$ equals to
\begin{equation}\label{spin}
p_{\ell}(Q) = \frac{1}{Z\,(2s+1)} \left\lbrace
\begin{split}
&\sum^{2s}_{k=\ell} e^{-\beta\omega(k-\ell)}, \hspace{0.65cm} 0 \leq \ell \leq 2s \\
  &\sum^{2s+\ell}_{k=0} e^{-\beta\omega(k-\ell)}, \hspace{0.4cm} -2s \leq \ell \leq 0
  \, .
\end{split}
\right.
\end{equation}
In this way, by explicitly computing the summations in Eq.\,(\ref{spin}), as well as the partition function $Z$, we obtain
\begin{equation}\label{eq:p_l_spin-s}
p_{\ell}(Q) = \frac{1}{\eta} \left\lbrace
\begin{split}
& 1 - e^{- \beta \omega (2s+1-\ell)}, \hspace{0.85cm} 0 \leq \ell \leq 2s \\
& e^{\beta \omega n} - e^{- \beta\omega (2s+1)}, \hspace{0.4cm} -2s \leq \ell \leq 0
\end{split} \right.
\end{equation}
with $\eta \equiv (1 - e^{- \beta\omega (2s+1)})(2s+1)$.
Eq.\,(\ref{eq:p_l_spin-s}) well reproduces the results
of the numerical simulations, as shown in Fig.\,\ref{fig:spin}.

\subsection{$\mathcal{O}$ and $H$ as quasi-commuting observables}

Now, for a generic spin-$s$ system with Hamiltonian $H = -\omega S_z$, let us consider the measurement observable $\mathcal{O} = \mathbf{\hat{n}} \cdot \mathbf{\hat{S}}$, with $\mathbf{\hat{n}} \equiv \sin\xi\,\mathbf{\hat{x}} + \cos\xi\,\mathbf{\hat{z}}$ and $\mathbf{\hat{S}} \equiv S_x\mathbf{\hat{x}} + S_y\mathbf{\hat{y}} + S_z\mathbf{\hat{z}}$.

On the one hand, it is worth noting that, if $\xi = 0$, then $[\mathcal{O},H] = 0$. Thus, by considering $\xi \ll 1$ (i.e., $[\mathcal{O},H]$ small), it holds that $\mathcal{O} = \xi S_x + S_z + O(\xi^2)$. On the other hand, we know that the eigenvalues of the spin operator $S_z$ are indexed by
$m \in \{ -s, -s+1,\ldots,s\}$ corresponding to the state vector $\ket{m}$. Hence, from the application of the first-order perturbation theory on the observable $\mathcal{O}$, we have that in the limit of small $\xi$ the eigenstates $\ket{\alpha_m}$ of $\mathcal{O}$ are equal to
\begin{equation}\label{eq:eigs_O_small_xi}
\ket{\alpha_m} = \ket{m} + \xi \sum_{m' \neq m} \frac{\bra{m'} S_x \ket{m}}{m-m'}  \ket{m'} + O(\xi^2).
\end{equation}
Since Eq.\,(\ref{eq:eigs_O_small_xi}) contains only the matrix elements of $S_x$ in the $S_z$-eigenbasis, it is now easy to compute the matrix $V$ up to higher order terms in $\xi$ by means of the expansion $V = e^{i\xi R} = \mathbb{I} + i\xi R + O(\xi^2)$. As a result, we find:
\begin{eqnarray}
R_{m,m'} &=& \frac{i}{2}\sqrt{(s-m)(s+m+1)} \delta_{m,m'+1}\nonumber \\
&-& \frac{i}{2}\sqrt{(s-m')(s+m'+1)} \delta_{m',m+1}\,.
\end{eqnarray}
In this way, concerning the transition matrix $L(\tau)$, the effective Hamiltonian $\Delta(\tau)$ (real symmetric operator) obeying Eq.\,(\ref{eq:L_small_xi}) is given by
$\Delta(\tau) = \mathcal{A} \sin^2 \frac{\omega \tau}{2}$, whose only non-zero elements are
\begin{equation}\label{eq:elements_op_A}
\begin{split}
\mathcal{A}_{m,m+1} &= - s(s+1) + m(m-1) \\
\mathcal{A}_{m,m-1} &= - s(s+1) + m(m+1) \\
\mathcal{A}_{m,m} &= 2(s(s+1)-m^2).
\end{split}
\end{equation}
As shown in Appendix A, the operator $\mathcal{A}$ can be diagonalized in the limit $s \gg 1$. The eigenvalues of $\mathcal{A}$ are equal to
\begin{equation}
a_k = k(k+1)\,,
\end{equation}
with $k=0,\ldots,2s$, while the $2s$ components $v_k(m)$ of the $k^{\text{th}}$ eigenvector $v_k$ are given by
\begin{equation}\label{eq:eigenvecs_mathcal_A}
v_k (m) = \sqrt{\frac{2k+1}{2s}} P_k \left( \frac{m}{s} \right)
\end{equation}
with $m=-s,-s+1,\ldots,s$. In Eq.\,(\ref{eq:eigenvecs_mathcal_A}), $P_k$ denotes the Legendre polynomial of order $k$.

This result suggests that, in the limit of $s \gg 1$,
the operator $\mathcal{A}$ can be expressed in terms of the orbital angular momentum $\mathbf{\hat{L}} \equiv L_x\mathbf{\hat{x}} + L_y\mathbf{\hat{y}} + L_z\mathbf{\hat{z}}$ of a single quantum particle.
By setting $\frac{m}{s} \equiv \cos\theta$, the eigenvalues and eigenstates of $\mathcal{A}$ coincide with the spectrum of $\mathbf{\hat{L}}^2$ provided that we limit ourselves to the sector $\mathcal{H}_A$ of the particle Hilbert space such that $L_z \mathcal{H}_A = 0$. Notice that the latter, in standard notation, corresponds to the part of the spectrum of $\mathcal{A}$ with $m=0$. This means that $\mathcal{A}$ can be written as
\begin{equation}
\mathcal{A} \simeq L_x^2 + L_y^2 + \mu L_z^2
\end{equation}
with $\mu \rightarrow \infty$. Under this limit, the euclidean evolution automatically excludes all the states that do not belong to $\mathcal{H}_A$.

\section{Large-N limit}\label{sec:large}

In this paragraph we determine analytical expressions describing the behaviour of a monitored quantum system in the limit of an infinite-dimensional Hilbert-space. Under this hypothesis, the theses of the Perron-Frobenius theorem are no longer valid \cite{Perron}, and, thus, it is no guaranteed that the largest eigenvalue $\lambda=1$ of $L(\tau)$ is non-degenerate.

For simplicity, let us take a spin-$s$ system, with Hamiltonian $H = - S_z/s$, and, as (intermediate) measurement observable, the Hermitian operator $\mathcal{O} = S_x$ (not commuting with the Hamiltonian). The scaling of the system Hamiltonian with $s$ has the usual purpose to maintain finite the range of the spectrum of $H$ as $s$ grows, and to help to retrieve the classical limit of an unit spin for $s\to\infty$.

Here, we are interested in predicting the thermalization of the analysed spin-$s$ system to the maximally mixed state, also in the limit of large-$s$ ($s\gg 1$, ideally infinite).

\begin{figure}[t!] \label{fig:collapse}
    \centering
    \includegraphics[scale=0.64]{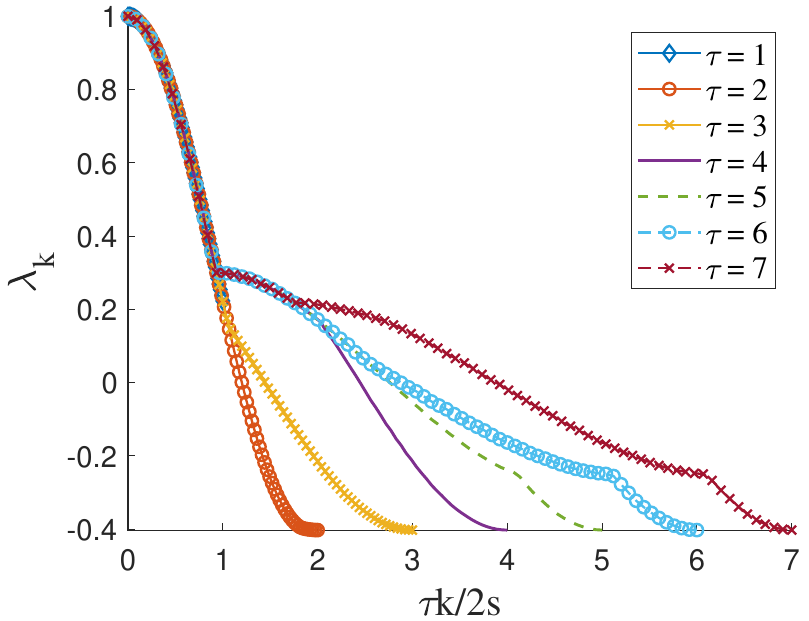}
    \caption{Comparison between the spectra of the stochastic matrices $L(\tau)$ with $s=300$ and different values of $\tau$, expressed in terms of the rescaled variable $\frac{k\tau}{2s}$. The plotted values of $\tau$ belongs to the set $\{1,2,\ldots,7\}$, and are respectively identified by the blue diamond, orange circle, yellow x-mark, and purple solid lines and green, cyan circle and red x-mark dotted lines. We can observe that, for $\frac{k\tau}{2s}$ smaller than a critical value (numerically determined to be $\approx 0.934$), all the data collapse on the same curve as predicted by Eq.\,(\ref{scaling}). In turn, for $\frac{k\tau}{2s} \ll 1$ we observe the quadratic behavior provided by Eq.\,\eqref{topspectrum}, namely $f(\frac{\tau k}{2s}) \approx 1-(\frac{\tau k}{2s})^2$.}
    \label{fig:spectrum}
\end{figure}

\begin{figure}
    \centering
    \subfloat{\includegraphics[scale=0.3]{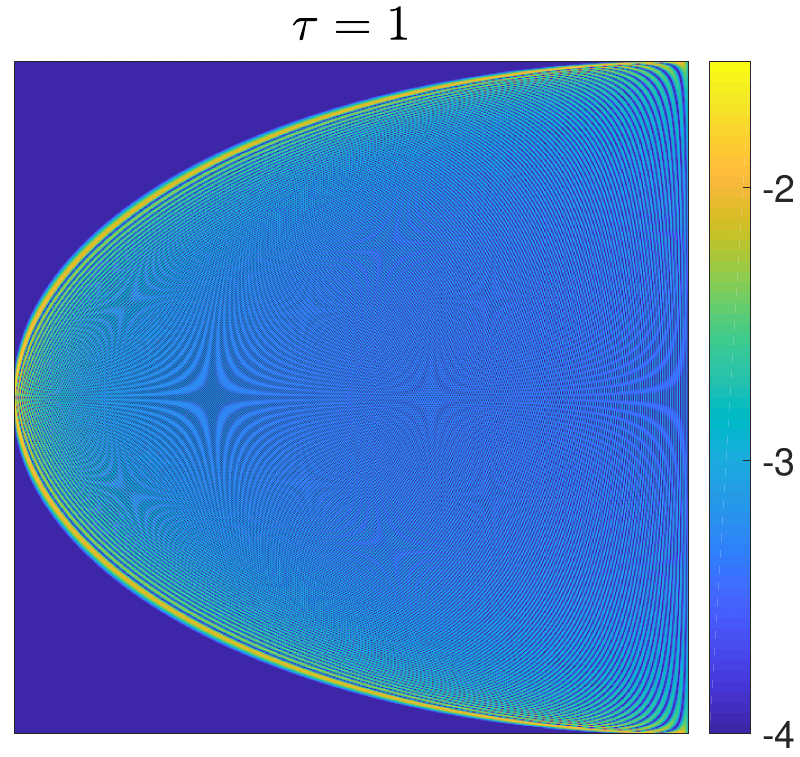}}
    \ \
    \subfloat{\includegraphics[scale=0.29]{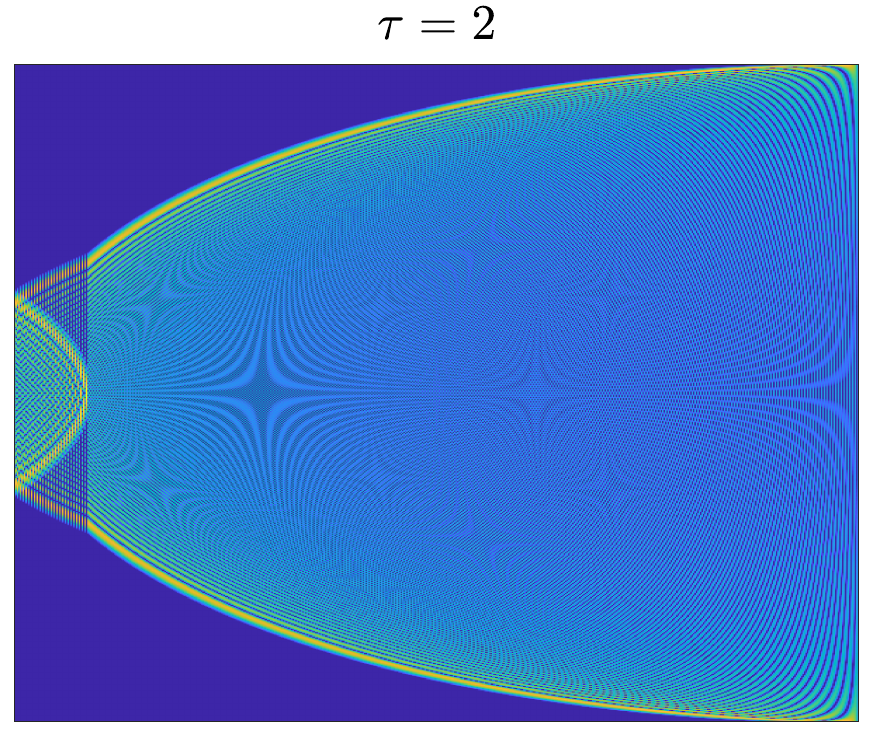}}
    \\
    \subfloat{\includegraphics[scale=0.285]{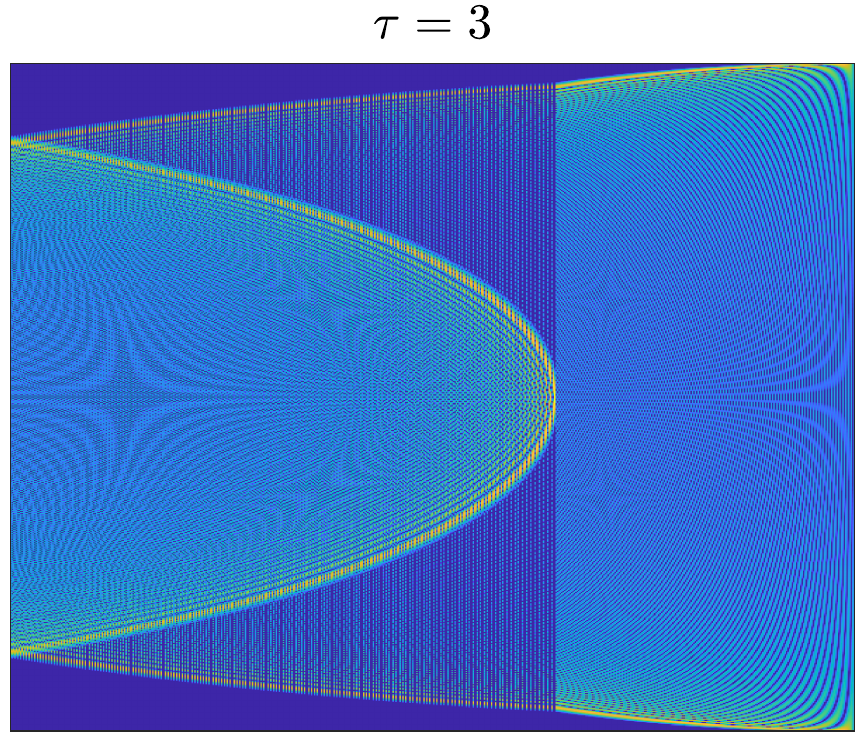}}
    \ \
    \subfloat{\includegraphics[scale=0.285]{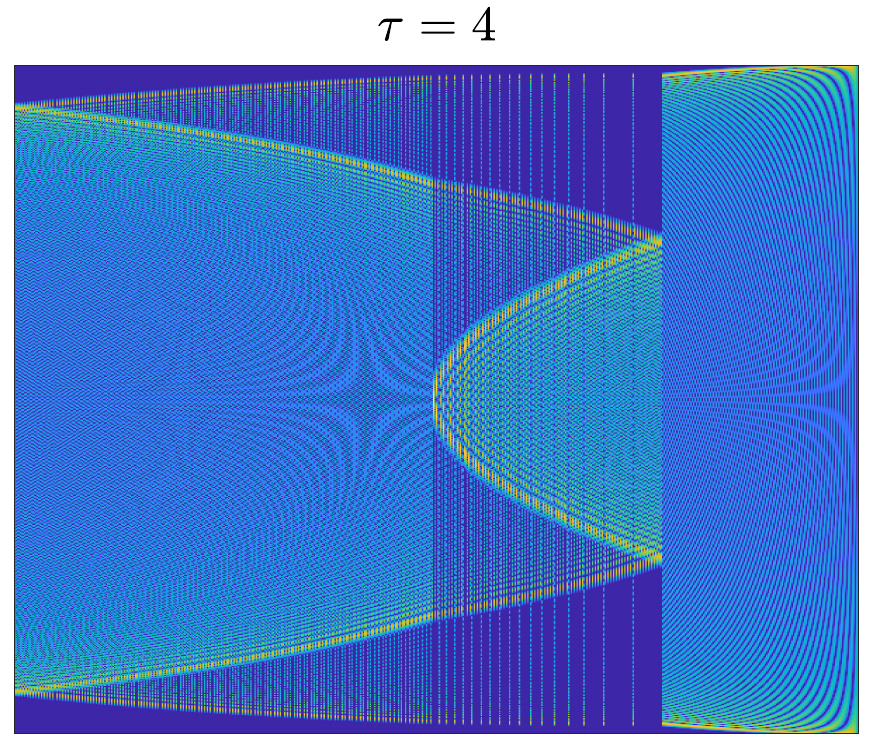}}
    \\
    \subfloat{\includegraphics[scale=0.285]{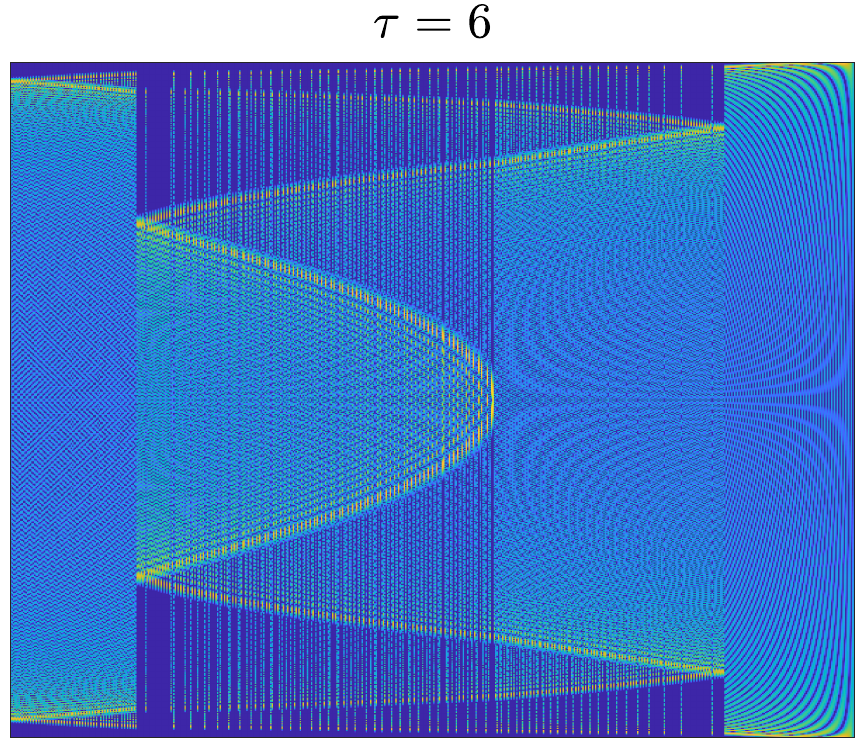}}
    \ \
    \subfloat{\includegraphics[scale=0.285]{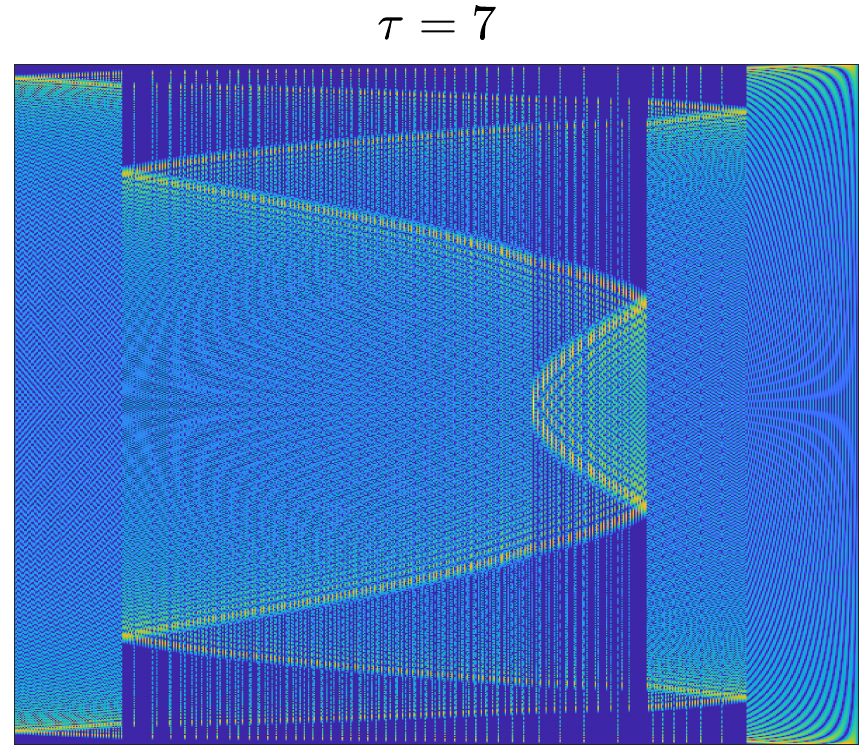}}
    \caption{Color plot of the matrix of eigenvectors of $L(\tau)$ for different values of $\tau$ (i.e., $\tau = 1,2,3,4,6,7$ from the top to the bottom and from left to right), with $s=300$. For visualization purposes, we plot on the y-axis of each panel the logarithm of each matrix element, while on the x-axis there is the index $k$ that labels each eigenvalue (the larger $k$, the larger the eigenvalue). We can observe that, in spite of the structures developed for the larger values of $\tau$, the eigenstates on the right of each panel, corresponding to the higher part of the spectrum, are practically the same as long as $\tau k/2s \lesssim 1$.}
    \label{fig:eigenvec}
\end{figure}

In Fig.\,\ref{fig:spectrum} we show the eigenvalues $\lambda_k$ of the transition matrix $L(\tau)$, with $\lambda_0 = 1 > \dots > \lambda_k > \dots > \lambda_{2s}$, for different choices of $\tau$. From the numerical simulations, we observe that the eigenvalues $\lambda_k$ tend to accumulate around $\lambda_0 = 1$. Moreover, it is also evident that, in the limit $s \gg 1$, the behavior of the highest eigenvalues is described by a universal function:
\begin{equation}\label{scaling}
    \lambda_k (\tau) \equiv f \left( \frac{\tau k}{2s} \right)
\end{equation}
with $f(0)=1$. This scaling relation is valid up to the critical value $\frac{\tau k}{2s}$ in correspondence of which a transition occurs. Notice that we have checked this evidence also for larger values of $s$. The critical value $\frac{\tau k}{2s}$ is found to be $\approx 0.934$ and it corresponds to the eigenvalue $\lambda_k(\tau) \approx 0.3$. As shown in Fig.\,\ref{fig:eigenvec}, a similar pattern is also present in the eigenvectors of the matrix $L(\tau)$. One can see that the eigenvectors, corresponding to small values of the index $k$ that labels them (so that $\frac{\tau k}{2s} \ll 1$), are independent on $\tau$.
Indeed, from Fig.\,\ref{fig:eigenvec}, one can observe that the the eigenvalues of the stochastic matrix $L(\tau)$, during their time evolution, behave as a propagating wave-front bouncing back and forth as time increases. In particular, each bounce is in correspondence of vertical lines -- identified by specific labels of the eigenvalues of $L(\tau)$ -- that moves closer and closer to the central label of the matrix, by maintaining ``frozen'' the eigenvalues with the largest value. Moreover, the time instants, in which a bounce occurs, correspond to a cusp in the eigenvalue distribution in Fig.\,\ref{fig:spectrum}. This evidence has been observed in the numerical simulations implemented for Figs.\,\ref{fig:spectrum} and \ref{fig:eigenvec}.

Independently of the nature of such transition, only the eigenvalues of $L(\tau)$ close to $1$ can affect the ITT. Thus, for our purposes, we will just focus on the spectrum of $L(\tau)$ that obeys to the scaling relation \eqref{scaling}, and we will analyze how the function $f$ behaves if its argument $\frac{\tau k}{2s}$ is small.

In doing this, let us consider the case $\tau \ll 1$ (for which of course $\frac{\tau k}{2s} \ll 1$). In this limit, the scaling relation \eqref{scaling} is valid for every $k=0,1,\ldots,2s$. Moreover, for small $\tau$, $\bra{\alpha_k} U (\tau) \ket{\alpha_\ell} = \delta_{k,\ell} - i \frac{\tau}{s} \bra{\alpha_k} S_z \ket{\alpha_\ell} + O(\tau^2)$,
so that for $k \neq \ell$
\begin{equation}
\bra{\alpha_k} L (\tau) \ket{\alpha_\ell} = \frac{\tau^2}{s^2} |\bra{\alpha_k} S_z \ket{\alpha_\ell}|^2 + O(\tau^3),
\end{equation}
while the diagonal elements are determined by imposing the constraint that $L(\tau)$ is a stochastic matrix. Thus, being $\{\ket{\alpha_k}\}$ the set of the eigenstates of $S_x$, we find that
\begin{equation} \label{LA}
L (\tau) = \mathbb{I} - \frac{\tau^2}{4s^2} \mathcal{A} + O(\tau^3),
\end{equation}
where $\mathcal{A}$ is the operator introduced in the previous paragraph and defined by Eq.\,(\ref{eq:elements_op_A}).

We conclude that in the limit $s \gg 1$ the spectrum of $L(\tau)$ is given by the eigenvalues
\begin{equation}
\lambda_k (\tau) = 1 -  \frac{\tau^2}{4 s^2} k(k+1) + O(\tau^3)
\end{equation}
with $k=0,\ldots,2s$. In this regard, it is worth noting that
$k(k+1) \approx k^2$ up to higher orders in $s^{-1}$ with the result that
\begin{equation}\label{topspectrum}
\lambda_k (\tau) = 1 - \left( \frac{\tau  k }{2s} \right)^2 + O(\tau^3)\,,
\end{equation}
in agreement with Eq.\,\eqref{scaling} for $f(x) = 1 - x^2 + O(x^3)$.
Accordingly, supported by our numerical analysis, we have that for large $s$ the value of the greatest eigenvalues of $L(\tau)$ (with $\frac{\tau k}{2s} \ll 1$) is correctly described by Eq.\,\eqref{topspectrum}, also for finite $\tau$. However, in the limit $M \gg 1$ only the eigenvalues close to $\lambda_0 = 1$ actually matter, since all the others are exponentially suppressed. Thus, when both $M$ and $s$ are large, one has that
\begin{equation}
L(\tau)^M \approx \left( \mathbb{I} - \left( \frac{\tau }{2 s} \right)^2 \mathcal{A} \right)^M \approx e^{- M\frac{\tau^2}{4 s^2}\mathcal{A}}\,.
\end{equation}
A different result is obtained depending on the order of
the limits $M \rightarrow \infty$ and $s \rightarrow \infty$. Indeed, if we perform the limit $M \rightarrow \infty$ while $s$ is finite, only the null eigenvector of $\mathcal{A}$ (corresponding to $k=0$ and $\lambda_0 = 1$) ``survives'' (is propagated over time without being nullify by a repeated sequence of products) and the system
thermalizes to an infinite-temperature state. Such finding is in accordance with our results obtained with a finite Hilbert space dimension and non-separable Hamiltonian. Conversely, by performing the limit $s \rightarrow \infty$ with $M$ finite, we get $L(\tau)^M
\rightarrow \mathbb{I}$. This means that the system becomes classical as $s \rightarrow \infty$, so that the measurements are no longer effective in changing the state of the system. Quite remarkably, also in this case, a non trivial result is obtained if we perform the two limits keeping $\frac{M \tau^2}{4 s^2} = \Tilde{t}$ constant. Indeed, one gets $L(\tau)^M \rightarrow e^{- \mathcal{A} \Tilde{t}}$ that corresponds again to a finite time Euclidean evolution with effective Hamiltonian $\mathcal{A}$, similarly to Eq.\,(\ref{eq:finite_time_Euclidean_ev}) for the case of $\mathcal{O}$ and $H$ quasi-commuting observables.

\section{Rotating two-dimensional gas}\label{sec:inc}

As example of partial ITT relevant for rotating two-dimensional gases \cite{Cooper}, let us consider a particle of mass $m\equiv 1$ moving in the $x$-$y$ plane and subjected to an anisotropic harmonic potential with frequencies $\omega_1 \neq \omega_2$, along the $x$ and $y$ directions respectively. Thus, the Hamiltonian is given by
\begin{eqnarray}\label{eq:H_two-dim_HO}
H &=& \frac{1}{2}\left(p_x^2 + p_y^2 + \omega_1^2 x^2 + \omega_2^2 y^2\right)\nonumber \\
&=& \omega_1 \left( a_x^{\dagger} a_x + \frac{1}{2} \right) + \omega_2  \left( a_y^{\dagger} a_y + \frac{1}{2} \right)
\end{eqnarray}
where $p_x$, $p_y$ denotes the momentum components of the particle in the
$x$, $y$ directions, and $a_x$, $a_y$ are the annihilation operators associated to the particle along $x$ and $y$. The energy eigenstates are given by $\ket{n_x,n_y}$ to which correspond the energy values $E = \omega_1 n_x + \omega_2 n_y$, being $\ket{n_x}$ and $\ket{n_y}$ the $1$D harmonic oscillator states along $x$ and $y$, respectively. As measurement observable $\mathcal{O}$, let us choose the pseudo-angular momentum
\begin{equation}
\widetilde{L} \equiv \frac{i}{2} \left( a_x^{\dagger} a_y - a_y^{\dagger} a_x \right) = \frac{1}{\sqrt{\omega_1 \omega_2}} (\omega_2 y p_x - \omega_1 x p_y).
\end{equation}
$\widetilde{L}$ is block diagonal on the eigenbasis of $H$. This can be seen by noting that $a_x a_y^{\dagger}\ket{n_x,n_y} \propto \ket{n_x -1,n_y+1}$ and $a^{\dagger}_x a_y \ket{n_x,n_y} \propto \ket{n_x +1,n_y-1}$. Thus, the action of $\widetilde{L}$ cannot generate any state with a different value of $n_x + n_y$. In other terms, each block with a given $n \equiv n_x + n_y$ is invariant under the action of the pseudo-angular momentum. Moreover, by computing the matrix elements of the pseudo-angular momentum, we can observe that, within each subspace with constant $n$, {\em (i)} $\widetilde{L}$ acts as (twice) the $y$-component of a spin-$s=n/2$ operator in the basis of the $z$-component, and {\em (ii)} $\widetilde{L}$ is not further reducible.

In conclusion, the thermalization process only involves the energy eigenstates $\ket{n_x,n_y}$ spanning a subspace with a fixed $n=n_x+n_y$, and the system behaves as a collection of independent spin-$s$ systems with $0<s<\infty$. Our findings are no longer valid if $\omega_1 = \omega_2 = \omega$. In such case, indeed, $\widetilde{L}$ becomes proportional to the angular momentum operator $\omega(yp_{x}-xp_{y})$ associated to an isotropic two-dimensional harmonic oscillator that commutes with $H$. Thus, no evolution is possible as well as ITT. It would be interesting to study the effect of repeated measurements of the pseudo-angular momentum in the slightly anisotropic case, with the aim to investigate in the interacting case whether and to what extent it could be usefully employed to reach quantum Hall states for two-dimensional rotating gases.

\section{Conclusions}\label{sec:concl}

In this paper, the asymptotic behaviour of a $N$-level quantum system subjected to a sequence of $M$ projective measurements is analyzed in the limit of $M$ and $N$ large. Such behaviour has been put in relation with common properties of the Hermitian operators $H$ (system Hamiltonian) and $\mathcal{O}$ (intermediate measurement observable), and peculiar characteristics of the heat distribution exchanged by the system with the external environment.

We have shown that, if $H$ and $\mathcal{O}$ do not share any common non trivial subspace, the final state of a monitored quantum system in the large-$M$ limit coincides with the maximally mixed state corresponding to a canonical thermal state with infinite temperature. We have denoted this latter condition as Infinite-Temperature Thermalization (ITT). Assuming the largest eigenvalues of the transition operator $L$ to be non-degenerate, we show how the ITT modifies the heat distribution associated to the monitored quantum system. Specifically, in the ITT regime, the initial and final energy outcomes, $\{E_n\}$ and $\{E_m\}$ respectively, are independent random variables and the corresponding characteristic function $G(u)$, with $u\in\mathbb{C}$, can be factorized in two distinct contributions just depending on the initial and final states.

Possible exceptions to ITT, to which we refer to as partial thermalization, can occur when the largest eigenvalue of the transition matrix operator $L$ is non longer non degenerate. Partial thermalization has been determined in the following three distinct cases. \\
{\em (i)} Whenever the Hermitian operators $H$ and $\mathcal{O}$ have one or more eigenvectors in common, as for example when $[H,\mathcal{O}]=0$. In such case, the ITT occurs only in partial way, since we no longer have the complete mixing of the intermediate measurement eigenvectors $|\alpha_k\rangle$, $k=1,\ldots,N$, at the end of the non-equilibrium quantum process. Indeed, what one can observe is the mixing of the eigenvectors $|v_r\rangle$ associated to the subspaces $S_r$ in which the Hamiltonian block matrices $H_r$ are defined. For the sake of clarity, we recall that the Hamiltonian blocks $H_r$ are the operators that compose the global Hamiltonian $H$ of the system, once expressed in the basis of $\mathcal{O}$. The presence of $R$ block matrices $H_r$ (and not just one) is the reason under the onset of a degeneracy of the eigenvalue $\lambda=1$ of $L(\tau)$, independently of the $\tau$-values. In this picture, the special case of $[\mathcal{O},H]=0$ is obtained for $R=N$. \\
{\em (ii)} ITT is not obtained when the value of the waiting times $\tau_j$ is on average much smaller than the inverse of the energy scale of the system, such that during
the application of two consecutive measurements the quantum system does not practically evolve and remains confined in its initial state. \\
{\em (iii)} Finally, analytical and numerical results in the large-$N$ limit, derived on a spin-$s$ system with $s\gg 1$, suggest that ITT can occur in the limit of $M \rightarrow \infty$ with $\tau \ll 1$ and a finite value of $s$. We found that the eigenvalues of $L(\tau)$ are the same for different values of $\tau$ as long as $\tau k/2s$ is smaller than a critical value that we estimated to be $\approx 0.934$. Interestingly, the matrix of eigenvectors displays a rich structure, but nevertheless the eigenstates corresponding to the larger eigenvalues are practically the same as soon as $\tau k/2 s \lesssim 1$, in agreement with the previously mentioned critical value. When, at variance, the limit $s \rightarrow \infty$ is taken with $M$ finite, we find that for $\tau \ll 1$ the application of a sequence of quantum projective measurements does not entail state changes within the measured quantum system, as one would expect in the classical limit.

As further remark, we would also like to stress that, experimentally, it is not necessary to perform an ideally infinite number $M$ of measurements to observe the theoretical results here exposed, even those valid in the asymptotic limit of $M$ large. In this regard, one could refer again to Ref.\,\cite{Hernandez2019} where a sequence of quantum projective measurements has been performed on a single nitrogen-vacancy (NV) center in diamond. In such experiments, indeed, a tendency of the quantum system towards an equilibrium thermal state with infinite temperature has been observed just after the application of less than $10$ projective measurements. We thus expect that this could be recovered also in other experimental platforms.

Our results are expected to pave the way for further investigations on monitored quantum systems, subjected to a sequence of non-projective quantum measurements\,\cite{WatanabePRE2014} and driven by time-dependent functions through Hamiltonian couplings. In such contexts, the distributions of both the heat and work, and their interplay according to the principles of thermodynamics, will have to be evaluated, e.g., by taking into account the cost of each applied projective measurement \cite{DeffnerPRE2016,GuryanovaQuantum2020}. Moreover, since we have adopted the more standard the two-point measurement scheme, it would be interesting to extend the obtained results by using different measurement schemes, such as the one recently explored in Refs.\,\cite{Sone20,Micadei20,G20,LevyPRXQ2020}. Finally, in light of the similarities between systems subjected to a repeated series of quantum measurements and periodically driven systems, one could investigate both continuous, single particle systems and many-body systems under repeated quantum measurements, possibly near a phase transition or in presence of an external dissipation\,\cite{RossiniPRB2020}, by using the results in this paper. In this respect, a very interesting example to be worked out in detail would be the Lipkin-Meshkov-Glick model, whose dynamical and entanglement properties have been intensively studied\,\cite{DefenuPRL18,Gabbrielli18SR}. In this regard, for such a model even the work and heat statistics and their relation with ground and also excited state quantum phase transitions have been recently addressed, as shown in Refs.\,\cite{WangPRE21,MzaoualiPRE21} and references therein. For the Lipkin-Meshkov-Glick model, the thermodynamical limit $N\to\infty$ and deviations from it was thoroughly investigated\,\cite{DusuelPRL05,LeyvrazPRL05,RibeiroPRE08,CanevaPRB08}; therefore, if subjected to a sequence of quantum measurements one could check (especially, in proximity of its quantum phase transitions) how the interplay between $N$ and the number $M$ of measurements arises, and whether the results obtained in this paper in the limit of large $N$ and $M$ apply.

\section*{Acknowlegments}
The authors gratefully acknowledge N. Fabbri, S. Hern\'andez-G\'omez and F. Poggiali for useful discussions. This work was financially supported by the MISTI Global Seed Funds MIT-FVG Collaboration Grant ``NV centers for the test of the Quantum Jarzynski Equality (NVQJE)'', and the MIUR-PRIN2017 project ``Coarse-grained description for non-equilibrium systems and transport phenomena (CO-NEST)'' No.\,201798CZL.

\section*{Appendix: Spectrum and eigenvectors of $\mathcal{A}$}

In this Appendix we to derive the spectrum and the eigenvectors of the operator $\mathcal{A}$. Let us start with the eigenvalues equation
\begin{equation}\label{eq1_appA}
\sum_{m'} \mathcal{A}_{m,m'} v(m') = a v(m) \,,
\end{equation}
equivalent to the relation
\begin{eqnarray}\label{eq2_appA}
a v(m) &=& 2(s(s+1) - m^2) v(m)\nonumber \\
&-& (s(s+1) - m(m+1)) v(m-1)\nonumber \\
&-& (s(s+1) - m(m-1)) v(m+1)
\end{eqnarray}
with $a$ and $v$ arbitrary eigenvalue and eigenvector of $\mathcal{A}$, respectively. Eq.\,(\ref{eq2_appA}) can be written as
\begin{eqnarray}\label{eq3_appA}
av(m) &=& (s(s+1) - m^2) ( 2v(m) - v(m+1) - v(m-1))\nonumber \\
&+& m (v(m+1) - v(m-1)).
\end{eqnarray}
In the limit $s \rightarrow \infty$, we assume that $v(m)$ is a smooth
function of the variable $x=\frac{m}{s} \in [-1,1]$. Thus, we make the
ansatz $v(m) = P(x)$ with $P(x)$ continuous function, so that
\begin{equation}\label{eq4_appA}
    v(m \pm 1) = P(x) \pm \frac{1}{s} P^{\prime}(x) + \frac{1}{2s^2}  P^{\prime \prime}(x)  + O(s^{-3}),
\end{equation}
where $P^{\prime}(x)$ and $P^{\prime \prime}(x)$ denote,
respectively, the first and second derivatives of $P(x)$ with respect to
$x$. As a result, the eigenvalue equation \eqref{eq3_appA}, up to $O(s^{-1})$ terms, is equal to
\begin{equation}\label{eq5_appA}
a P(x) = - \frac{1}{s^2} (s(s+1) - s^2 x^2) P^{\prime \prime} (x) + \frac{2sx}{s} P^{\prime} (x)
\end{equation}
whereby, by taking the limit $s \rightarrow \infty$, we finally get
\begin{equation}\label{eq6_appA}
(1 - x^2) P^{\prime \prime} (x) - 2x P^{\prime} (x) + a P(x) = 0\,.
\end{equation}
Eq.\,\eqref{eq6_appA} is the well-known Legendre equation. In order to
have normalizable solutions of the Legendre equation in the interval
$x \in [-1,1]$, one has to set that the eigenvalue $a$ belongs to the
set $\{a_k\}$ with $a_k=k(k+1)$ and $k$ integer $\geq 0$. Thus, in this case, the eigenfunctions are proportional to the $k$-order Legendre polynomials $P_k(x)$. In conclusion, by enforcing the normalization condition, we find:
\begin{equation}\label{eq7_appA}
v_k(m) = \sqrt{\frac{2k+1}{2s}} P_k \left( \frac{m}{s} \right)
\end{equation}
where the variable $s$ at the denominator of the normalization factor is required to pass from the normalization in $x$ to that in $m$.

\end{document}